\documentclass[aps,prl,reprint,showpacs]{revtex4-1}
\usepackage[dvipdfmx]{graphicx}

\usepackage{amsmath,amssymb}
\usepackage{color}
\usepackage{siunitx}

\usepackage{bm}
\def\vb#1{{\bm#1}}
\def\v#1{\mathbf{#1}}			

\def\vq{\v{q}} 					% def. of vector "q"
\def\vk{\v{k}} 					% def. of vector "k"
\def\vI{\v{I}}
\def\vS{\v{v}_{\vb{\sigma}}}

\def\vS{\v{S}}

\def\r{\v{r}} 					% def. of vector "r"
\def\q{\v{q}} 					% def. of vector "q"
\def\k{\v{k}} 					% def. of vector "k"
\def\vsigma{\vb{\sigma}}

\def\del{\partial}

\def\la{\langle}
\def\ra{\rangle}

\def\hac{h_{\rm ac}}
\def\Jsd{J_{\rm sd}}

\def\RE{{\rm Re}} % real part
\def\IM{{\rm Im}} % imaginary part
\def\ikqom{\int_{\vq\vk\omega}} % integral by q, k, and omega
\def\TR{{\rm Tr}}%trace

\begin{document}

%\preprint{}

\title{Nuclear spin pumping by pulling effect}



\author{Y. Ohnuma$^{1,2}$, S. Maekawa$^{2,1}$, and M. Matsuo$^{1,2}$}
%\email{mamoru@ucas.ac.cn}
\affiliation{%
${^1}$Kavli Institute for Theoretical Sciences, University of Chinese Academy of Sciences, Beijing, 100190, China.\\
${^2}$RIKEN Center for Emergent Matter Science (CEMS), Wako, Saitama 351-0198, Japan.
}%

\date{\today}

\begin{abstract}
The nuclear-to-electron spin angular momentum conversion via hyperfine coupling in a normal metal (NM)/ferromagnet (FM) bilayer system is theoretically investigated by using the nonequilibrium Green's function method. 
The spin current generated by the nuclear magnetic resonance (NMR) is found to be enhanced by the pulling effect in the FM when the temperature is lower than NMR resonance frequency. 
In a Co/Pt bilayer system, we show that the spin current by NMR becomes larger than that of the ferromagnetic resonance (FMR). 
\end{abstract}

%72.25.-b	Spin polarized transport (for spin polarized transport devices, see 85.75.-d)
%75.76.+j	Spin transport effects (for devices exploiting spin polarized transport, see 85.75.Hh, 85.75.Mm, and 85.75.Ss)
%47.10.ad Navier-Stokes equations
%47.61.Fg	Flows in micro-electromechanical systems (MEMS) and nano-electromechanical systems (NEMS)

\maketitle %\maketitle must follow title, authors, abstract and \pacs

%%%%%%%%%%%%%%%%%%%%%%%%%%%%%%%%%%%%%%%%%%%%%%%%%%%%%%%%%%%%%%%%%%%%%%%
%%%%%%%%%%%%%%%%%%%%%%%%%%%%%%%%%%%%%%%%%%%%%%%%%%%%%%%%%%%%%%%%%%%%%%%
\paragraph{Introduction.---}
Spin current, a flow of electron spins, is a key concept in the field of spintronics~\cite{MaekawaEd2012}.  Generation of spin current has been demonstrated by using angular momentum conversion between spin and various angular momenta in condensed matter such as magnetization\cite{SaitohE:APL88:2006}, photons\cite{PierceDT:PRB13:1976, AndoK:APL96:2010}, the orbital motion of electrons, and mechanical angular momentum carried by moving materials\cite{TakahashiR:NP12:2016, MatsuoM:PRB87:2013, KobayashiD:PRL119:2017}. 
In this context, a remaining angular momentum in condensed matter is nuclear spins. 

Because a nuclear spin couples to an electron spin via the hyperfine coupling, the nuclear spin can excite the nonequilibrium electron spin dynamics, and then, generate a spin current in principle. 
However, 
the interconversion of nuclear spin and electron spin has not been exploited.
One of the reasons for this situation is that the hyperfine coupling between electron and nuclear spins is rather small compared to the couplings among electron spins. In addition, the time scale of nuclear spin is much slower than that of electron spin.   

In order to overcome this difficulty, we consider the amplification of the spin current using the pulling effect~\cite{deGennes1963}. As noted above, the modulation of electron spins by the motion of the nuclear spins is negligible due to the mismatch of their resonance frequencies. In ferromagnets with a large density of nuclei at low temperature, however, the dynamics of electron spins is modulated by the nuclear spins because the coherent motion of high density of the nuclear spins is induced. At the nuclear magnetic resonance (NMR) condition, the electron spins adiabatically follow the nuclear spins. This effect is called the pulling effect, and in this case, the spin angular momentum of the nuclei transfers to the electrons with high efficiency since the nuclear and electron spins behave as a coupled system. Hence, the pulling effect is expected to amplify the spin current generated by the nuclear spins.

In this article, we theoretically investigate the spin-current generation due to interconversion of nuclear spins into electron spins via the hyperfine coupling in a normal metal (NM)/ferromagnet (FM) bilayer system. 
We formulate a spin transport theory driven by the nuclear spin dynamics in FM using the nonequilibrium Green's function method. 
It is shown that the spin current generated at the interface is enhanced by the pulling effect and takes the maximum value when the temperature is lower than NMR frequency. 
In a Co/Pt bilayer system, it is shown that the spin current generated by NMR at low temperature is larger than that of the ferromagnetic resonance (FMR). Our theory provides a new method of generating the spin current using NMR (MHz frequency range) larger than that generated by FMR (GHz frequency range).

\vspace{1cm}
\begin{figure}[!tb]
\begin{center}
\includegraphics[scale=0.5]{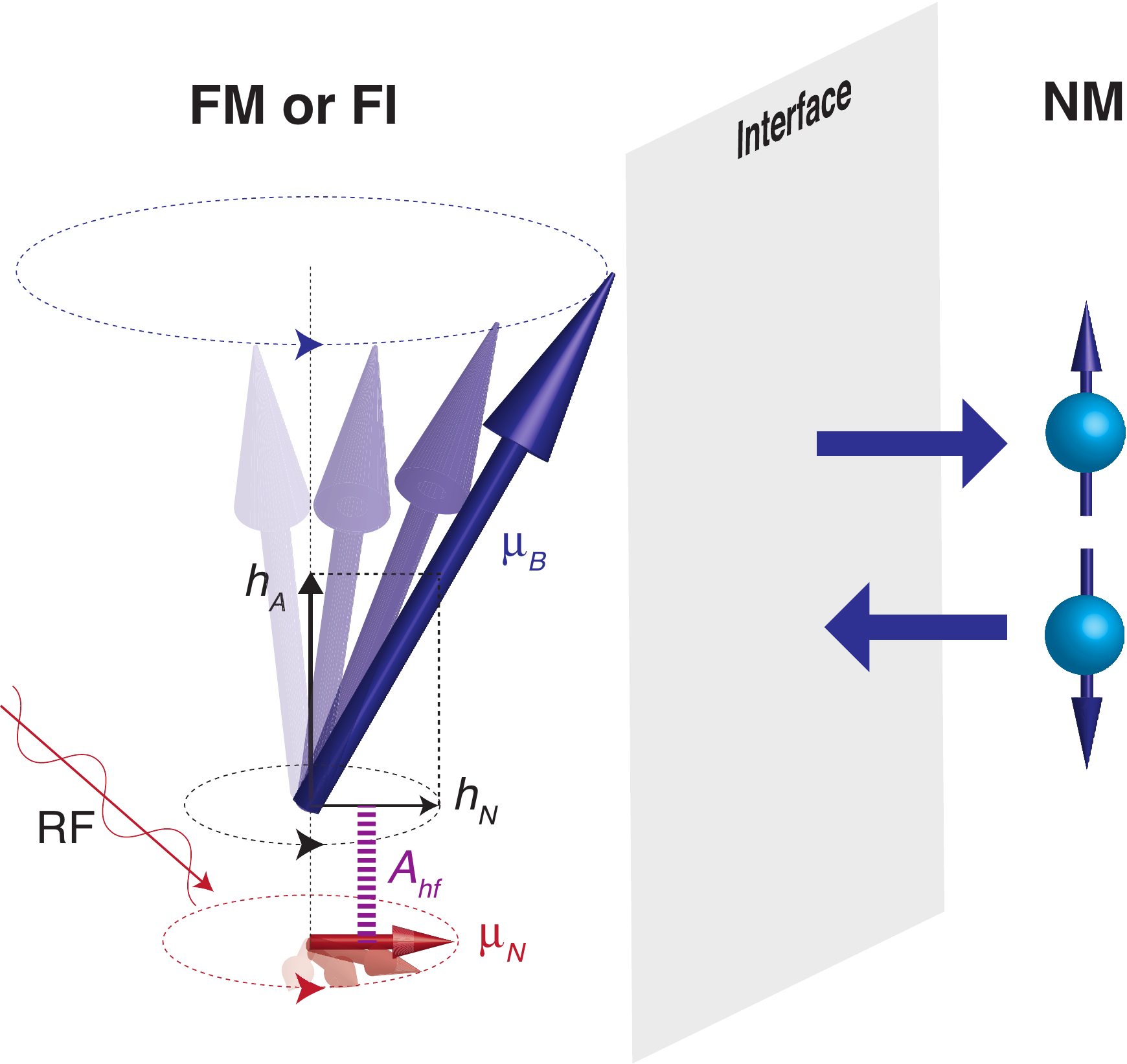}
\caption{(Color online) Schematic illustration of spin pumping driven by NMR via pulling effect. Here, $\mu_B$, $h_A$ and $h_N$ are the Bohr magneton, anisotropic field and hyperfine field acting on electron spins, respectively, and $A_{hf}$ and $\mu_N$ are coupling constant of hyperfine interaction and the nuclear magnetic moment, respectively.}
\label{fig_1}
\end{center}
\end{figure}

\paragraph{Model.---}
We consider the spin transport in a bilayer system, where a normal metal (NM) and a ferromagnet (FM) are coupled to each other through the $s$-$d$ exchange at the interface:
\begin{eqnarray}
H_{\rm int} = \Jsd \sum_i \vsigma_i \cdot \vS_i,
\end{eqnarray}
where $\vsigma_i$ and $\vS_i$ are conduction spin in NM and localized electron spin in FM on the i-th site at the interface, and $\Jsd$ is the exchange coupling. 
In addition, localized spins in FM are coupled to nuclear spins via hyperfine coupling:
\begin{eqnarray}
H_{\rm IS} =A_{hf} \sum_j \vI_j \cdot \vS_j, 
\end{eqnarray}
where $\vI_j$ is nuclear spin on the i-th site and $A_{hf}$ is hyperfine coupling constant.  

The spin current generated at the interface is given by the rate of change of conduction electron spin in the NM, $\mathcal{I}_S : =\la \hat{\mathcal{I}}_S \ra = \hbar \sum_i \la \del_t \sigma^z_i \ra $, where $\hat{\mathcal{I}}_S := \hbar \sum_i  \del_t \sigma^z_i $ is a spin current operator and $\la \cdots \ra := \TR [ \hat{\rho} \cdots ]$ denotes the statistical average with the density matrix $\hat{\rho}$. 
By performing the second-order perturbation  with respect to the interfacial exchange coupling, the generated spin current is given by:
\begin{eqnarray}
\mathcal{I}_S 
  &=&  \frac{\Jsd^2N_{\rm int}}{\hbar^2} \RE\!\! \ikqom \!\! [ \chi^{R}_{\q\r,\omega t} G^{<}_{\k\r', \omega t} + \chi^{<}_{\q\r, \omega t}G^{A}_{\k\r',\omega t} ], 
\end{eqnarray}
where $N_{\rm int}$ is the number of sites at the interface and the random average is taken over the impurity positions at the interface.   
The lesser (retarded) Green's function for conduction electron spin, $\chi_{\q\r,\omega t}^{<(R)}$, is defined as 
$\chi_{\q\r,\omega t}^{<(R)} := \int \exp[i\k \cdot \delta \r-i \omega \delta t]\chi^{<(R)}(\r+\delta r\, t+\delta t,\r-\delta \r \, t-\delta t)$, 
$\chi^{<}(\r_1t_1,\r_2t_2) = -i\la \sigma^-_{\r_2t_2}\sigma^+_{\r_1t_1}\ra $, 
$\chi^{R}(\r_1t_1,\r_2t_2) = -i\theta(t_1-t_2)\la [\sigma^+_{\r_1t_1},\sigma^-_{\r_2t_2}]\ra $, 
and $\theta(t)$ is the step function. 
The lesser (advanced) Green's function for localized spin, $G_{\k\r',\omega t}^{<(A)}$, is defined by 
$G_{\k\r',\omega t}^{<(A)} := \int \exp[i\k \cdot \delta \r-i \omega \delta t] G^{<(A)}(\r+\delta r\, t+\delta t,\r-\delta \r \, t-\delta t)$, 
$G^{<}(\r_1t_1,\r_2t_2) = -i\la S^-_{\r_2t_2}S^+_{\r_1t_1} \ra $, and 
$G^{A}(\r_1t_1,\r_2t_2) = i\theta(t_2-t_1)\la [S^+_{\r_1t_1},S^-_{\r_2t_2}]\ra $. 
$\chi^R_{\q\omega}$ is given by~\cite{Fulde1968} $\chi^R_{\q\omega} = \chi_N\tau_{\textrm{sf}}(1+\lambda^2_{N}\q^2+i\omega\tau_{\textrm{sf}})^{-1}$, with $\chi_N$, $\tau_{\textrm{sf}}$, and $\lambda_N$ being the paramagnetic susceptibility, the spin-flip relaxation time, and the spin-diffusion length in NM, respectively.

Let us consider the situation, where conduction electron spins in the NM are in local thermal equilibrium whereas 
localized spins in the FM are excited by nuclear spins via the hyperfine coupling, describing by $\delta G^<$. The spin current is reduced to 
\begin{eqnarray}
\mathcal{I}_S  = \frac{\Jsd^2N_{\rm int}}{\hbar^2}\int_{\q\k\omega} \IM \chi_{\q\omega}^R \IM \delta G_{\k\omega}^<. \label{IS}
\end{eqnarray}

\paragraph{Spin pumping by pulling effect.---}
From now on, 
we calculate the lesser function of localized electron spin excited by the pulling effect\citep{deGennes1963}.
A coupled system of localized electron and nuclear spins is modeled by the following Hamiltonian: $H_{\rm FM}=  H_{\rm S} + H_{\rm I} + H_{\rm ac}+ H_{\rm IS}$, where 
\begin{eqnarray}
&&H_{\rm S} \! = J\sum_{\la i,j\ra}\bm{S}_i\cdot\bm{S}_j + \hbar \gamma_e h_0 \sum_j S_j^z + \frac{D}{2} \sum_j (S_j^z)^2, \\
&&H_{\rm I} \! =  -\hbar \gamma_N h_0 \sum_j I_j^z \!\! \\
&&H_{\rm ac} = - \hbar\gamma_N \hac \sum_j (I_j^x \!\cos \nu t \!+\! I_j^y \!\sin \nu t),
\end{eqnarray}
where $J$ is the exchange coupling with $\sum_{\la i,j\ra}$ being the summation over nearest-neighbor sites, $\gamma_e$ is the electron gyromagnetic ratio, $D$ is magnetic anisotropy constant,
$\gamma_N$ is the nuclear gyromagnetic ratio, 
$h_0$ is a DC external magnetic field, and 
$\hac$ and $\nu$ are the amplitude and frequency of the AC magnetic field, respectively.

The dynamics of the local electron and nuclear spins are given by the Landau-Lifshitz-Gilbert (LLG) equation and the Bloch equation:
\begin{eqnarray}
&&\dot{\vS} = \gamma_e (\vS \times \v{h}_e) + \frac{\alpha}{S_0} \vS \times \dot{\vS}, \\
&& \dot{I}^{x,y} \!\! = \! \gamma_N (\vI \times \v{h}_I)^{x,y} \!\! - \!\frac{I^{x,y}\!\!}{T_2}, \dot{I}^z \!\! = \! \gamma_N (\vI \times \v{h}_I)^z \!\! - \! \frac{I^z_0 \!\! -\!\! I^z\!\!}{T_1},
%\\
\end{eqnarray}
where 
$\alpha$ is the Gilbert damping constant of the FI, $\v{h}_e$ and $\v{h}_I$ are the magnetic fields acting on localized spin and nuclear spins, and $T_1$ and $T_2$ are the longitudinal and transverse relaxation times of nuclear spins, respectively.

The magnetic field acting on the $j$-site localized spin is calculated by
$h_{e,j}^{a} =  \frac{\partial (H_{\rm I} + H_{\rm S})}{\hbar \gamma_e \partial S_j^a}$ with $a = x,y,z$, and thus,  
\begin{eqnarray}
&&h_{e,j}^{x,y} = \frac{A_{hf}}{\hbar \gamma_e} I_j^{x,y},
h_{e,j}^{z} = h_0 + h_A + \frac{A_{hf}}{\hbar \gamma_e} I_j^{z},
\end{eqnarray}
where $h_A = D\la S^z \ra $ is the anisotropic field. Here, we introduce the thermal and site averaged z-component of electron spins $\la S^z \ra $ given by $\la S^z \ra:=(N_e)^{-1}\sum^{N_e}_j \la S^z_j \ra$ with $N_e$ being the total number of sites of electron spins. 

Because the localized spin dynamics is much faster than the nuclear spin dynamics, 
$|\gamma_N/\gamma_e |\ll 1$, the localized spins adiabatically follow the nuclear spins in FM. In this case, the localized spin can be considered to be static: $\dot{\vS} \approx 0$.
Then, the transverse component of localized magnetic moment is related to the longitudinal one as $S^{x,y} = (h_e^{x,y}/h_e^z)S^z$, and we obtain 
\begin{eqnarray}
S_j^{\pm} = \frac{h_N}{h_0+h_A+ h_N\la I^z\ra/\la S^z \ra} I_j^{\pm}.
\label{Sjxy1}
\end{eqnarray}
Here, we replace approximately the $j$ dependent z component of the nuclear spins $I_j^z$ by the thermal and site averaged values as $\la I^z \ra$ given by $\la I^z \ra:=(N_I)^{-1}\sum^{N_I}_j \la I^z_j \ra$ with $N_I$ being the total number of sites of nuclear spins. In the above equation $S_j^{\pm}$ and $I_j^{\pm}$ are given by $S_j^{\pm}=S_j^x \pm S_j^y$ and $I_j^{\pm}=I_j^x \pm I_j^y$, respectively, and the hyperfine field $h_N$ is defined as $h_N := A_{hf}\la S\ra \//(\hbar \gamma_e)$

Similarly, the magnetic field $h_I$ is calculated by 
$h_{I,j}^{a} =  \frac{\partial (H_{\rm I} + H_{\rm IS})}{\hbar \gamma_N \partial I_j^a}$:
\begin{eqnarray}
&&h_j^\pm (t) = \hac e^{\pm i \nu t} - \frac{A_{hf}}{\hbar \gamma_N}S_j^\pm, h_j^z = h_0  - \frac{A_{hf}}{\hbar \gamma_N}S_j^z, 
\end{eqnarray}
where $h_j^\pm = h_j^x \pm h_j^y$. 
Using these relations, the Bloch equation can be rewritten as
\begin{eqnarray}
\frac{d}{dt}I^\pm_j = \pm i(\tilde{\nu}_N I^\pm_j + \gamma_N\hac\la I^z\ra e^{\pm i \nu t}) -\frac{I^\pm_j}{T_2},\label{Itpm}
\end{eqnarray}
where $\tilde{\nu}_N$ is the modified NMR frequency given by $\tilde{\nu}_N = \gamma_N h_0 + \nu_N (1+\xi\la I^z\ra/\la S^z\ra)$ with $\nu_N$ and $\xi$ being the bare NMR frequency and enhancement factor defined as $\nu_N:=A_{hf}\la S^z\ra/\hbar$ and   $\xi:=S^{\pm}_j/I^{\pm}_j$, respectively. 
Inserting $I^\pm(t) = \int^{\infty}_{-\infty} d\nu/2\pi I^\pm_\nu e^{i\nu t}$ into Eq. (\ref{Itpm}), we have
\begin{eqnarray}
I^\pm_\nu = \mp\frac{\la I^z \ra \gamma_N h_{ac}e^{\pm i\nu t}}{\nu \pm \tilde{\nu}_N +iT_2^{-1}}.\label{Inupm}
\end{eqnarray}
Using Eqs.~(\ref{Sjxy1}) and (\ref{Inupm}), the lesser Green's function $\delta G^<$ is given by
\begin{eqnarray}
\delta G^<_{\k \omega} = -i  \Big|  \frac{ \xi \gamma_N \hac\la I^z \ra}{\nu + \tilde{\nu}_N + i T_2^{-1}} \Big|^2 \delta (\omega - \nu). \label{deltaG<}
\end{eqnarray}
By inserting this equation into Eq. (\ref{IS}), we obtain the spin current generated by nuclear spin dynamics: 
\begin{eqnarray}
\mathcal{I}^{\textrm{Pull}}_S = \frac{\Jsd^2 N_{\rm int}}{\hbar^2} \int_{\q} \IM\chi^R_{\q\nu}\frac{ (\xi \gamma_N \hac\la I^z \ra ) ^2}{ (\nu + \tilde{\nu}_N)^2 + (1/T_2)^2}. \label{IS-Iz}
\end{eqnarray}

At the resonance condition $\nu=-\gamma_N \tilde{h}_z$, Eq.~(\ref{IS-Iz}) reduces to  
\begin{eqnarray}
\mathcal{I}^{\textrm{Pull}}_S = &&-G_s(\tilde{\nu}_N)\mathcal{A}_{\rm int}(\la I^z \ra \xi)^2 \tilde{\nu}_N (T_2)^2( \gamma_N \hac)^2 
\label{IS-Iz2}
\end{eqnarray}
where $\mathcal{A}_{\rm int}$ is the surface area of the interface expressed by $\mathcal{A}_{\rm int}=N_{\rm int} a_{\rm int}$ with $a_{\rm int}$ being the unit surface area of the interface, and $G_s(\tilde{\nu}_N)$ is given by $G_s(\tilde{\nu}_N)=(\Jsd^2/\hbar) a^{-1}_{\rm int}\int_{\q} \IM\chi^R_{\q\tilde{\nu}_N}/\tilde{\nu}_N$.

Equation (\ref{IS-Iz2}) shows that NMR spin pumping is proportional to the square of the transverse relaxation time $T_2$. Because the transverse components of nuclear spins relax to the thermal equilibrium state during $T_2$, the long $T_2$ leads to the strong non-equilibrium state and enhances the nuclear spin pumping. 

The temperature dependence of the nuclear spin pumping is determined mainly by $(\la I^z \ra \xi)^2$ in Eq.~(\ref{IS-Iz2}). Here, we calculate the z-component of nuclear spin $\la I^z \ra$ in mean field approximation given by
\begin{eqnarray}
\la I^z \ra \!= \! I_0 \Big[ \frac{2I_0+1}{2I_0} \coth\!\Big( \frac{2I_0+1}{2} x\Big)\!-\!\frac{1}{2I_0}\coth\frac{x}{2} \Big],
\label{Eq:Iz}
\end{eqnarray}
where $I_0$ is nuclear spin value and $x$ is defined as $x:=\hbar\tilde{\nu}_z/(k_B T)$. 
When the temperature $T$ is lower than $T^*$ given by $T^*:=\hbar\tilde{\nu}_N/k_B$, the nuclear spins are fully polarized and $\la I^z\ra$ becomes $I_0$. 
Because the factor $(\la I^z \ra \xi)^2$ is an increasing function of $\la I^z \ra$, NMR spin pumping is enhanced for $T\ll T^*$. 
By contrast, when the temperature $T$ is higher than $T^*$, the temperature and magnetic field dependence of NMR spin current is obtained as $\mathcal{I}^{\textrm{Pull}}_S \propto (|h_N|-h_0)^3/h_0{}^2T^2$. 

\paragraph{Amplification of NMR spin pumping.---}
\vspace{1cm}
\begin{figure}[!tb]
\begin{center}
\includegraphics[scale=0.25]{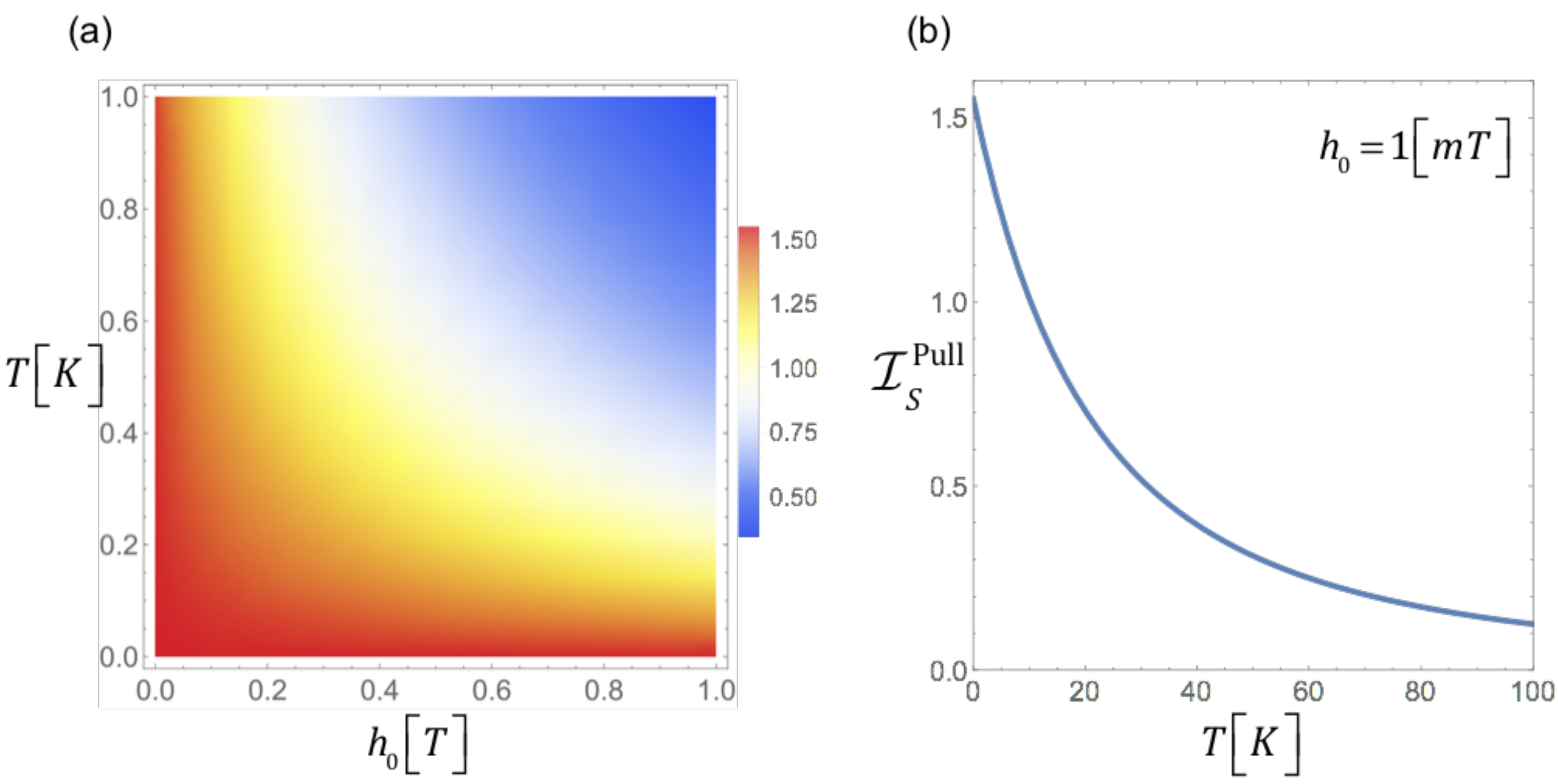}
\caption{(Color online) (a) Spin current signal $\mathcal{I}_S^{\textrm{Pull}}$ plotted as a function of $T$ and $h_0$ for a Co/Pt bilayer system. The plotted spin current is scaled by FMR spin pumping $\mathcal{I}_S^{\textrm{FMR}}=1.2\times10^7$~A/m$^2$. (b) Temperature dependence of $\mathcal{I}_S^{\textrm{Pull}}$ at a fixed magnetic field $h_0=1$~mT. }
\label{fig_TH}
\end{center}
\end{figure}

Now we estimate the spin current~(\ref{IS-Iz2}) for a bilayer system of the cobalt and platinum (Co/Pt) where Co and Pt are FM and NM layers, respectively. To evaluate Eq.~(\ref{IS-Iz2}), we combine Eq.~(\ref{IS-Iz2}) with the spin current driven by FMR. Following Ref.~\onlinecite{Ohnuma2014}, we obtain the spin current driven by FMR as follows:
\begin{eqnarray}
\mathcal{I}^{\textrm{FMR}}_S= G_s(\omega_0)\mathcal{A}_{\rm int}\la S^z \ra^2\frac{(\gamma_e h_{ac})^2}{\alpha^2\omega_0},
\label{Eq:FMR-Is}
\end{eqnarray}
where $\alpha$ is the Gilbert damping constant given by $\alpha=\alpha_0+\delta\alpha$, with $\alpha_0$ and $\delta\alpha$ being the intrinsic and additional terms due to the spin pumping, and $\omega_0$ is FMR frequency expressed by $\omega_0=\gamma_e(h_0+h_A)$. 
We introduce $G_s(\omega_0)$ as $G_s(\omega_0):=(\Jsd^2/\hbar) a^{-1}_{\rm int}\int_{\q} \IM\chi^R_{\q\omega_0}/\omega_0$. 
%we obtain $G_s(\omega_0)\approx G_s(\nu_N)$. 

Using the material parameters in a Co/Pt system\cite{Rojas2014} as $\alpha=0.014$, $h_{\rm ac}=0.11$~mT, $\gamma_e h_0=2\pi\times9.75$~GHz 
%, the relaxation time $\tau_0$ of platinum estimated as $\tau_0\approx5$~fs, 
and $\mathcal{I}^{\textrm{FMR}}_S=1.2\times10^{7}$~A/m$^2$, we obtain %$G_s(\tilde{\nu}_N)=1.7\times10^{18}$~m$^{-2}$ and 
$\Jsd^2\chi_N/a_{\textrm{int}}\approx5.2\times10^{17}$~eV/m$^2$. 
Combining these parameters and the parameters for $^{59}$Co in the fcc cobalt, $\gamma_N=6.3015$~kHz/Oe, $ \la S^z \ra= 0.85$\cite{Kawakami72}, $I_0=7/2$
, $h_A=125$~Oe at 3 K\cite{deGennes1963,Bromer78}, $h_N\approx20$~T, $T_2=20\times10^{-6}$~s 
~\cite{Enokiya76}, $T^*\approx10$~mK and the spin current generated from NMR at $T=10$~mK and $h_0=1$~mT is $\mathcal{I}^{\textrm{Pull}}_S=1.5\times10^{15}\times h^2_{\rm ac}$~A/(T$\cdot$m$^2$). 

Substituting $h_{\rm ac}=0.11$~mT into this result, we show the temperature and magnetic field dependence of nuclear spin pumping $\mathcal{I}^{\textrm{Pull}}_S$ for a Co/Pt system in Fig.~\ref{fig_TH}(a), where the spin current is normalized by FMR spin pumping $\mathcal{I}^{\textrm{FMR}}_S$. In the region of low temperature or magnetic field (red colored region), the spin current driven by NMR becomes larger than that of FMR. In \ref{fig_TH}(b), we show the temperature dependence of $\mathcal{I}^{\textrm{Pull}}_S$ at a fixed magnetic field $h_0=1$~mT. The NMR spin pumping is a decreasing function of the temperature. 
We obtain the NMR spin current as $|\mathcal{I}^{\textrm{Pull}}_S|=1.9\times10^7$~A/m$^2$ 
at $T=10$~mK and $h_0=1$~mT.
Comparing $|\mathcal{I}^{\textrm{Pull}}_S|$ with the FMR spin pumping in a Co/Pt system $\mathcal{I}^{\textrm{FMR}}_S$~\cite{Rojas2014}, the NMR spin pumping at $T=10$~mK and $h_0=1$~mT is amplified to about $1.5$ times 
larger than that of the FMR spin pumping. Note that FMR spin pumping in a Co/Pt system is almost independent of temperature~\cite{Verhagen16}. 

The enhancement of NMR spin pumping in a Co/Pt system at low temperature is obtained the competition of the gyromagnetic ratios and relaxation times of the nuclear and electron spins. 
Comparing Eqs.~(\ref{IS-Iz2}) with (\ref{Eq:FMR-Is}), we obtain the ratio of $\mathcal{I}^{\textrm{Pull}}_S$ to $\mathcal{I}^{\textrm{FMR}}_S$ as $\mathcal{I}^{\textrm{Pull}}_S/\mathcal{I}^{\textrm{FMR}}_S\approx(\xi\la I^z\ra/\la S^z\ra)^2 (\gamma_N/\gamma_e)^2 \alpha  (T_2/T_{\textrm{FMR}})$, where we introduce the relaxation time of FMR $T_{\textrm{FMR}}$ as $T_{\textrm{FMR}}=[(\alpha_0+\delta\alpha)\omega_0]^{-1}$. Considering $T_{\textrm{FMR}}$ calculated as $T_{\textrm{FMR}}\approx 10^{-8}$~s, we find $\mathcal{I}^{\textrm{Pull}}_S/\mathcal{I}^{\textrm{FMR}}_S\approx C(\xi\la I^z\ra/\la S^z\ra)^2$ with $C$ being the numerical constant of the order of 1. It is expected that the ratio of $\mathcal{I}^{\textrm{Pull}}_S$ to $\mathcal{I}^{\textrm{FMR}}_S$ is enhanced in FM with large $\alpha$ and long $T_2$. 

\paragraph{Conclusion.---}
In this article, we have investigated spin-current generation by nuclear spin dynamics via hyperfine coupling in a normal metal (NM)/ferromagnet (FM) bilayer system. 
We have formulated spin transport theory using the nonequilibrium Green's function method. 
The spin current generated at the interface is found to be enhanced by the pulling effect and is maximized at the temperature lower than NMR resonance frequency. 
In a Co/Pt system, we have predicted the amplification of the NMR spin current generation.
Our theory reveals a new mechanism of angular momentum conversion in condensed matter systems, and 
suggests a new method of generating the spin current using NMR (MHz frequency range) which is larger than that generated by FMR (GHz frequency range).

\paragraph{Acknowledgements.---}
The authors thank H. Chudo, M. Imai and K. Yamamoto for valuable discussions. 
This work is financially supported by ERATO-JST (JPMJER1402), and KAKENHI (No. 26103005, No. JP16H04023, and No. JP26247063) from MEXT, Japan.


\end{document}